\documentclass[twoside]{article}
\usepackage{amsmath, amsfonts, amssymb}
\usepackage{epsf}
\usepackage{graphicx}
\usepackage{aat}
\usepackage{tabularx}

\begin{document}

\def\apj{ApJ}
\def\aj{AJ}
\def\apjl{ApJL}
\def\apjs{ApJS}
\def\mnras{MNRAS}
\def\aaps{A\&AS}
\def\aap{A\&A}
\def\apss{Astrophysics and Space Science}
\def\araa{ARA\&A}
\def\pasp{PASP}

\thispagestyle{myfirst}

\setcounter{page}{1}

\mytitle{The intrinsic shape of bulges} \myauthor{J. M\'endez-Abreu$^{1,2}$, E. Simonneau$^{3}$, J. A. L. Aguerri$^{1,2}$, E. M. Corsini$^{4}$}
\myadress{$^{1}$Instituto Astrof\'\i sico de Canarias, E-38200 La Laguna, Spain\\
         $^{2}$Departamento de Astrof\'\i sica, Universidad La Laguna, E-38205 La Laguna, Spain \\ 
         $^{3}$Institut d'Astrophysique de Paris, C.N.R.S.-U.P.M.C., F-75014 Paris, France\\
         $^{4}$Dipartimento di Astronomia, Universit\`a di Padova, I-35122 Padova, Italy
}
\mydate{(Received December , 2010)}
\myabstract
{The $J$-band  structural parameters  of a magnitude-limited  sample of
148 unbarred  S0--Sb galaxies were derived using  the GASP2D algorithm
and then analyzed  to derive the intrinsic shape  of their bulges.  We
developed a new  method to derive the intrinsic  shape of bulges based
only on photometric data  and on the geometrical relationships between
the apparent and  intrinsic shapes of bulges and  disks. The method is
conceived as  completely independent of the studied  class of objects,
and it  can be applied whenever  a triaxial ellipsoid  embedded in (or
embedding) an axisymmetric component is considered
We found that the intrinsic  shape is well constrained for a subsample
of 115 bulges with favorable  viewing angles. A large fraction of them
is characterized by an  elliptical section ($B/A<0.9$).  This fraction
is $33\%$, $55\%$, and $43\%$  if using their maximum, mean, or median
equatorial ellipticity, respectively. Most of them are flattened along
their polar axis ($C<(A+B)/2$).
The distribution  of triaxiality is strongly  bimodal. This bimodality
is driven by  bulges with S\'ersic index $n >  2$, or equivalently, by
the bulges of galaxies with a bulge-to-total ratio $B/T > 0.3$. Bulges
with $n \leq 2$ and with $B/T \leq 0.3$ follow a similar distribution,
which is different  from that of bulges  with $n > 2$ and  with $B/T >
0.3$. In particular, bulges with $n\leq2$ and with $B/T \leq 0.3$ show
a  larger fraction  of  oblate axisymmetric  (or nearly  axisymmetric)
bulges,  a smaller  fraction  of triaxial  bulges,  and fewer  prolate
axisymmetric (or  nearly axisymmetric)  bulges with respect  to bulges
with $n > 2$ and with $B/T > 0.3$, respectively.
According  to  predictions  of  the  numerical  simulations  of  bulge
formation,  bulges with  $n \leq  2$, which  show a  high  fraction of
oblate axisymmetric (or nearly axisymmetric) shapes and have $B/T \leq
0.3$, could be  the result of dissipational minor  mergers. Both major
dissipational  and  dissipationless mergers  seem  to  be required  to
explain the variety of shapes found for bulges with $n > 2$ and $B/T >
0.3$.
}

\mykey{galaxies: bulges -- galaxies: photometry --
  galaxies: statistics -- galaxies: structure}

\section{Introduction}

The  intrinsic  shapes of  elliptical  galaxies  and  disks have  been
extensively studied  in the literature.  However, bulges appear  to be
less studied,  even if  they account for  about $25\%$ of  the stellar
mass of the local universe \cite{driver07}.

The  study of  the intrinsic  shape of  bulges  presents similarities,
advantages, and drawbacks with respect to that of elliptical galaxies.
For  bulges, the  problem  is  complicated by  the  presence of  other
luminous components and requires the accurate isolation of their light
distribution.  On  the other hand,  the presence of the  galactic disk
allows for the  accurate constraining of the inclination  of the bulge
under  the assumption  that the  two components  share the  same polar
axis.

Although the kinematical properties  of many bulges are well described
by  dynamical  models of  oblate  ellipsoids  which  are flattened  by
rotation  with  little  or  no  anisotropy  \cite{daviesillingworth83,
  corsini99,  pignatelli01},  the  twisting  of  the  bulge  isophotes
\cite{zaritskylo86} and the misalignment between the major axes of the
bulge  and disk  \cite{bertola91, mendezabreu08}  observed  in several
galaxies are not possible if the bulge and disk are both axisymmetric.
These features were interpreted as the signature of bulge triaxiality.
This  idea is  also  supported  by the  presence  of non-circular  gas
motions  \cite{gerhardvietri86,  falconbarroso06,  pizzella08}  and  a
velocity  gradient  along   the  galaxy  minor  axis  \cite{corsini03,
  coccato04, coccato05}.
Perfect  axisymmetry is  also ruled  out when  the intrinsic  shape of
bulges is  determined by statistical analyses based  on their observed
ellipticities.   Bertola et al.   \cite{bertola91} measured  the bulge
ellipticity and the  misalignment between the major axes  of the bulge
and  disk in  32 S0--Sb  galaxies. They  found that  these  bulges are
triaxial  with  mean  axial  ratios  $\langle  B/A  \rangle=0.86$  and
$\langle C/A \rangle=0.65$. In  contrast, measurements of $\langle B/A
\rangle=0.79$  for  the bulges  of  35  early-type  disk galaxies  and
$\langle B/A \rangle=0.71$ for the bulges of 35 late-type spirals were
found by Fathi \& Peletier \cite{fathipeletier03}.  None of the 21 disk
galaxies  with  morphological types  between  S0  and  Sab studied  by
Noordermeer \& van der Hulst \cite{noordermeervanderhulst07} harbors a
truly  spherical bulge.  They  obtain a  mean flattening  $\langle C/A
\rangle=0.55$.  Mosenkov  et al.  \cite{mosenkov10}  obtained a median
value of  the flattening  $\langle C/A \rangle=0.63$  for a  sample of
both early and late-type edge-on galaxies in the near infrared.

In M\'endez-Abreu et al. \cite{mendezabreu08}(Paper I) we measured the
structural    parameters    of    a    sample    of    148    unbarred
early-to-intermediate  spiral galaxies using  the GASP2D  algorithm to
analyze  their  near-infrared  surface-brightness  distribution.   The
probability  distribution  function  (PDF)  of  the  bulge  equatorial
ellipticity   was   derived  from   the   distributions  of   observed
ellipticities of bulges and misalignments between bulges and disks. We
proved  that about  $80\%$ of  the sample  bulges are  not  oblate but
triaxial  ellipsoids with  a mean  axial ratio  $\langle  B/A\rangle =
0.85$.

In this  work, (see  M\'endez-Abreu et al.   \cite{mendezabreu10}
for details) we  introduce a new method to  derive the intrinsic shape
of  bulges  under the  assumption  of  triaxiality.  This  statistical
analysis is  based upon the analytical relations  between the observed
and intrinsic shapes  of bulges and their surrounding  disks and it is
applied to  the galaxy sample described  in Paper I.   The method make
use only of photometric data  and have been conceived to be completely
independent of  the studied  class of objects,  and it can  be applied
whenever   triaxial   ellipsoids  embedded   in   (or  embedding)   an
axisymmetric component are considered.

\section{Basic geometrical considerations}

We  assume that  the bulge  is a  triaxial ellipsoid  and the  disk is
circular and  lies in  the equatorial plane  of the bulge.   Bulge and
disk share the same center  and polar axis. Therefore, the inclination
of  the polar  axis (i.e.,  the galaxy  inclination) and  the position
angle of  the line of  nodes (i.e., the  position angle of  the galaxy
major  axis) are directly  derived from  the observed  ellipticity and
orientation of the disk, respectively.

Let ($x,  y, z$) be the  Cartesian coordinates with the  origin in the
galaxy  center,  the  $x-$axis   and  $y-$axis  corresponding  to  the
principal equatorial axes of the bulge, and the $z-$axis corresponding
to the polar axis of the bulge  and disk. If $A$, $B$, and $C$ are the
lengths of the ellipsoid semi-axes (without considering $A \geq B \geq
C$), the equation of the bulge in its own reference system is given by

\begin{equation} 
\frac{x^2}{A^2} + \frac{y^2}{B^2} + \frac{z^2}{C^2} = 1. 
\label{eqn:ellipsoid} 
\end{equation} 

\begin{figure*}[!t]
  \centering
  \includegraphics[width=\textwidth]{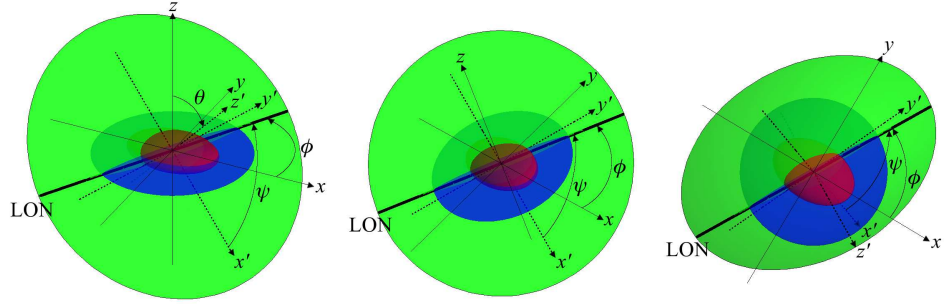}
  \caption{Schematic three-dimensional view of the ellipsoid geometry.
    The bulge ellipsoid,  the disk plane, and the  sky plane are shown
    in red,  blue, and green, respectively.  The  reference systems of
    the ellipsoid and  of the observer as well as  the LON are plotted
    with thin solid lines, thin  dashed lines, and a thick solid line,
    respectively.   The  bulge ellipsoid  is  shown  as  seen from  an
    arbitrary  viewing  angle (left  panel),  along  the LOS  (central
    panel), and along the polar axis (right panel).}
  \label{fig:geom}
  \end{figure*}

Let  $(x',y',z')$ be  now the  Cartesian coordinates  of  the observer
system. It  has its origin in  the galaxy center,  the polar $z'-$axis
along  the line of  sight (LOS)  and pointing  toward the  galaxy. The
plane of  the sky lies on  the $(x',y')$ plane. The  projection of the
disk onto the sky plane is an  ellipse whose major axis is the line of
nodes  (LON), i.e.,  the  intersection between  the  galactic and  sky
planes.  The  angle  $\theta$   between  the  $z-$axis  and  $z'-$axis
corresponds  to the  inclination of  the galaxy  and therefore  of the
bulge ellipsoid; it can be derived as $\theta=\arccos{(d/c)}$ from the
length $c$  and $d$ of the  two semi-axes of the  projected ellipse of
the disk.
We defined $\phi$ ($0 \leq \phi \leq \pi /2$) as the angle between the
$x$-axis and the LON on the equatorial plane of the bulge $(x,y)$.
Finally, we  also defined $\psi$  ($0 \leq \psi  \leq \pi /2$)  as the
angle  between   the  $x'$-axis   and  the  LON   on  the   sky  plane
$(x',y')$. The three angles $\theta$, $\phi$, and $\psi$ are the usual
Euler  angles  and  relate  the  reference  system  $(x,y,z)$  of  the
ellipsoid with  that $(x',y',z')$  of the observer  by means  of three
rotations (Fig. \ref{fig:geom}). Indeed, because  of the location
of  the LON  is  known, we  can  choose the  $x'-$axis  along it,  and
consequently it holds that $\psi=0$.
Now,  if we  identify the  latter with  the ellipse  projected  by the
observed ellipsoidal bulge, we can  determine the position of its axes
of symmetry $x_{\rm e}$ and $y_{\rm e}$ and the lengths $a$ and $b$ of
the  corresponding semi-axes.  The  $x_{\rm e}-$  axis forms  an angle
$\delta$  with the  LON  corresponding  to the  $x'-$axis  of the  sky
plane. We always choose $0 \le  \delta \le \pi/2$ such that $a$ can be
either the major  or the minor semi-axis.  Later  we will explain that
this riddle is solved because  the two possibilities coincide, and one
is the mirror image of the other.

Following \cite{simonneau98}, we are able to express the length of the
bulge semi-axes ($A$, $B$, and $C$) as a function of the length of the
semi-axes  of the  projected ellipse  ($a$, $b$)  and the  twist angle
($\delta$).
 
\begin{small} 
\begin{eqnarray} 
A^2 & = & K^2\left( 1 + \frac{e\,\sin2\delta}{1 + e\,\cos2\delta}\frac{\tan\phi}{\cos\theta}\right) ,  \label{eqn:A} \\ 
B^2 & = & K^2\left( 1 - \frac{e\,\sin2\delta}{1 + e\,\cos2\delta}\frac{\cot{\phi}}{\cos\theta}\right) , \label{eqn:B} \\ 
C^2 & = & K^2\left( 1 - \frac{2\,e\,\cos2\delta}{\sin^2\theta\left(1 + e\,\cos2\delta\right)} + \frac{2\,e\,\cos\theta\,\sin2\delta}{\sin^2\theta\left(1 + e\,\cos2\delta\right)} \cot^2{\phi} \right) \label{eqn:C}. 
\end{eqnarray} 
\end{small} 

where $K^2 = \frac{a^2+b^2}{2}\left[ 1 + e\,\cos 2\delta \right]$, and
$e=\frac{a^2 -  b^2}{a^2 + b^2}$.  The  values of $a,  b, \delta,$ and
$\theta$ can  be directly obtained  from observations.  Unfortunately,
the  relation  between  the  intrinsic and  projected  variables  also
depends  on the spatial  position of  the bulge  (i.e., on  the $\phi$
angle), which is  actually the unique unknown of  our problem. Indeed,
this will constitute the basis of our statistical analysis.

\subsection{Characteristic angles} 
\label{sec:angles} 

There  are physical  constraints which  limit the  possible  values of
$\phi$,  such as the  positive length  of the  three semi-axes  of the
ellipsoid \cite{simonneau98}. Therefore, we define some characteristic
angles   which  constrain   the  range   of  $\phi$.    Two  different
possibilities must be taken into account for any value of the observed
variables $a$, $b$, $\delta$ and $\theta$.

The first case corresponds to $a>b$.  It implies that $e>0$ and $A>B$.
For  any value of  $\phi$, $A^2>K^2$  and $K^2$  is always  a positive
value.  On the  other hand, $B^2$ and $C^2$ can  be either positive or
negative depending  on the value of  $\phi$. This limits  the range of
the values of $\phi$.  $B^2$ is positive only for $\phi> \phi_B$.  The
angle $\phi_B$ is defined by $B^2 = 0$ in Eq. \ref{eqn:B} as

\begin{equation}
\tan{\phi_B}= \frac{e\,\sin{2\delta}}{\cos{\theta}\,\left( 1+e\,\cos{2\delta}\right)}.
\label{eqn:B=0}
\end{equation}

Likewise,  $C^2$ is  positive only  for values  of  $\phi<\phi_C$. The
angle $\phi_C$ is defined by $C^2=0$ in Eq. \ref{eqn:C} as

\begin{equation}
\tan{2\phi_C}= \frac{2\,e\,\sin{2\delta}\,\cos{\theta}}{e\,\cos{2\delta}\,\left(1+\cos^2{\theta}\right)-\sin^2{\theta}}.
\label{eqn:C=0}
\end{equation}

Thus, if  $a>b$ then  the values of  $\phi$ can  only be in  the range
$\phi_B \le \phi \le \phi_C$.

The  second case  corresponds to  $a<b$.   It implies  that $e<0$  and
$A<B$.   For any  value of  $\phi$, $B^2>K^2$  and $K^2$  is  always a
positive  value.  But, $A^2$  and  $C^2$  can  be either  positive  or
negative depending on  the value of $\phi$.  This  limits the range of
the values  of $\phi$.   $A^2$ is positive  only for $\phi  < \phi_A$.
The angle $\phi_A$ is defined by $A^2 = 0$ in Eq. \ref{eqn:A} as

\begin{equation}
\tan{\phi_A}= -\frac{\cos{\theta}\left(1+e\,\cos{2\delta}\right)}{e\,\sin{2\delta}}.
\label{eqn:A=0}
\end{equation}

Likewise,  $C^2$ is  positive only  for values  of  $\phi>\phi_C$. The
angle $\phi_C$  is given in  Eq. \ref{eqn:C=0}.  Thus, if  $a<b$, then
the  values $\phi$  can only  be  in the  range $\phi_C  \le \phi  \le
\phi_A$.

However, the  problem is  symmetric: the second  case, when  the first
semi-axis of  the observed ellipse corresponds to the minor  axis (i.e., $a<b$),  is the mirror
situation of the first case,  when the first measured semi-axis of the
observed ellipse corresponds  to the major axis (i.e.,  $a>b$). In the
second  case, if we  assume the  angle $\pi/2  -\delta$ to  define the
position  of the  major semi-axis  $a$  of the  observed ellipse  with
respect to the  LON in the sky plane, and $\pi/2  -\phi$ to define the
position of the  major semi-axis $A$ of the  equatorial ellipse of the
bulge with respect  to the LON in the bulge  equatorial plane, then we
can always consider  $a>b$ and $A>B$.  Therefore, $e\ge  0$ and $E \ge
0$ always.  This means that  we have the same mathematical description
in both cases: the possible values  of $\phi$ are $\phi_B \le \phi \le
\phi_C$ with $\phi_B$ and  $\phi_C$ defined by Eqs.  \ref{eqn:B=0} and
\ref{eqn:C=0},  respectively. 
According to this definition oblate and prolate triaxial ellipsoids do
not necessarily have an axisymmetric shape.  A detailed description of
all these cases will be given at the end of this section.

We define the  quadratic mean radius of the  equatorial ellipse of the
bulge   in   order   to   extensively  discuss   all   the   different
possibilities.

\begin{equation}
R^2 = \frac{A^2+B^2}{2} = K^2\tan{\phi_B}\left[\cot{\phi_B} - \cot{2\,\phi}\right],
\label{eqn:R}
\end{equation}
%
which depends only on the unknown position $\phi$.

Since $A^2>B^2$,  $A^2 \ge R^2  \ge B^2$ but  there is always  a value
$\phi_{\rm RC}$ corresponding to the case $C^2=R^2$

\begin{equation}
\tan{2\phi_{RC}}= \tan{2\delta}\,\frac{1+\cos^2{\theta}}{2\,\cos{\theta}}.
\end{equation}

\begin{figure*}[!t]
  \centering
  \includegraphics[width=\textwidth,height=7cm]{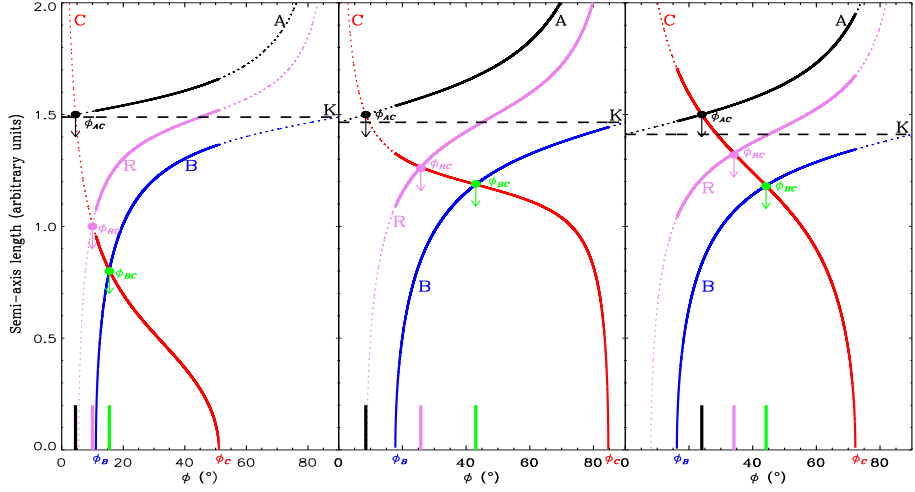}
  \caption{The lengths $A$, $B$, and $C$ of the semi-axes of the bulge
    ellipsoid and its mean equatorial  radius $R$ as a function of the
    angle  $\phi$.   The  solid  lines  correspond to  the  ranges  of
    physically possible  values of $A$,  $B$, $C$, and $R$,  while the
    dotted  lines show  their overall  trends within  $0 \le  \phi \le
    \pi/2$. A  triaxial bulge with  $\phi_{AC} < \phi_{RC}  < \phi_B$,
    $\phi_{AC}  <  \phi_B <  \phi_{RC}$,  and  $\phi_B  < \phi_{AC}  <
    \phi_{RC}$  is  shown  in  the  left, central,  and  right  panel,
    respectively.}
  \label{fig:angles}
  \end{figure*}

The mean equatorial radius allows us to distinguish oblate ($C^2<R^2$)
and  prolate  ($C^2> R^2$)  triaxial  ellipsoids.  Unfortunately,  the
situation  is   more  complicated  due  to  the   existence  of  other
intermediate angles  such as  $\phi_{BC}$ which occurs  when $B^2=C^2$
and $\phi_{AC}$ which occurs  when $A^2=C^2$. In particular, there are
four  different possibilities  for the  intrinsic shape  of  the bulge
ellipsoid.   They are sketched  in Fig.   \ref{fig:angles} and  can be
described as follows

\begin{itemize}

\item If  $\phi_{AC} < \phi_{RC}  < \phi_B$ the triaxial  ellipsoid is
  always  oblate (Fig.  \ref{fig:angles},  left panel).  It is  either
  completely  oblate  (i.e.,  $A>B>C$)  if $R>B>C$  ($\phi_{BC}<  \phi
  <\phi_C$)  or  partially  oblate   if  $R>C>B$  ($\phi_B  <  \phi  <
  \phi_{BC}$).

\item If $\phi_{AC}  < \phi_B < \phi_{RC}$ the  triaxial ellipsoid can
  be  either  oblate   or  prolate  (Fig.   \ref{fig:angles},  central
  panel). It is either  completely oblate if $R>B>C$ ($\phi_{BC}< \phi
  <  \phi_C$), or  partially  oblate if  $R>C>B$  ($\phi_{RC}< \phi  <
  \phi_{BC}$),  or partially  prolate  if $C>R>B$  ($\phi_B  < \phi  <
  \phi_{RC}$).

\item If $\phi_B < \phi_{AC} < \phi_{RC}$ four different possibilities
  are  allowed   for  the  triaxial  shape  of   the  bulge  ellipsoid
  (Fig. \ref{fig:angles}, right panel). It is either completely oblate
  if  $R>B>C$ ($\phi_{BC}<  \phi <  \phi_C$), or  partially  oblate if
  $R>C>B$ ($\phi_{RC}  < \phi <  \phi_{BC}$), or partially  prolate if
  $A>C>R$  ($\phi_{AC} <  \phi  < \phi_{BC}$),  or completely  prolate
  (i.e., $C>A>B$) if $C>A>R$ ($\phi_B < \phi < \phi_{AC}$).
  
\end{itemize}
\section{Intrinsic shape of bulges}
\label{sec:3d}

The strength of this new method  is that now we are able to constraint
the shape  of bulges  within well defined  range of axis  ratios, and,
moreover,  since   the  axis  ratios   are  given  by   a  probability
distribution  function, a  complete  3D dimensional  picture of  their
shape can be obtained.  The equation linking the intrinsic ellipticity
and flattening axis ratio is given by

\begin{scriptsize}
\begin{eqnarray}
\frac{2\,\sin{\left(2\phi_C\right)}}{F_{\rm \theta}}\frac{C^2}{A^2} =
\sin{\left(2\phi_C-\phi_B\right)} \sqrt{\left(1-\frac{B^2}{A^2}\right)^2 - \sin^2{\phi_B}\left(1+\frac{B^2}{A^2}\right)^2} -
\sin{\phi_B}\cos{\left(2\phi_C - \phi_B\right)}\left(1+\frac{B^2}{A^2}\right)^2.
\label{eqn:vartie}
\end{eqnarray}
\end{scriptsize}

which constrain the intrinsic shape of an observed bulge with the help
of the known characteristic angles $\phi_B$ and $\phi_C$, which depend
only   on   the  measured   values   of   $a$,   $b$,  $\delta$,   and
$\theta$.

\begin{figure*}[!t]
  \centering
  \includegraphics[width=0.49\textwidth]{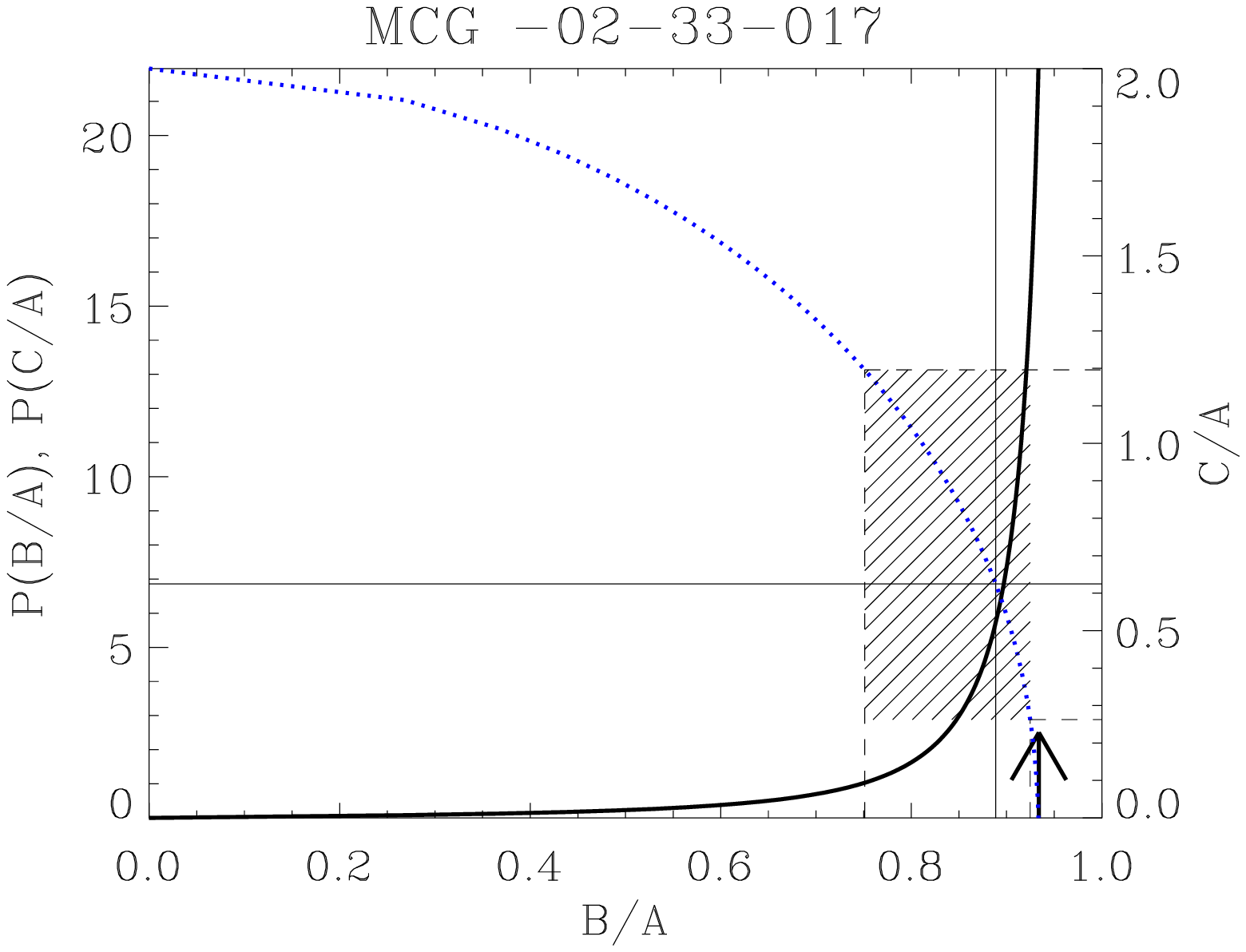}
  \includegraphics[width=0.49\textwidth]{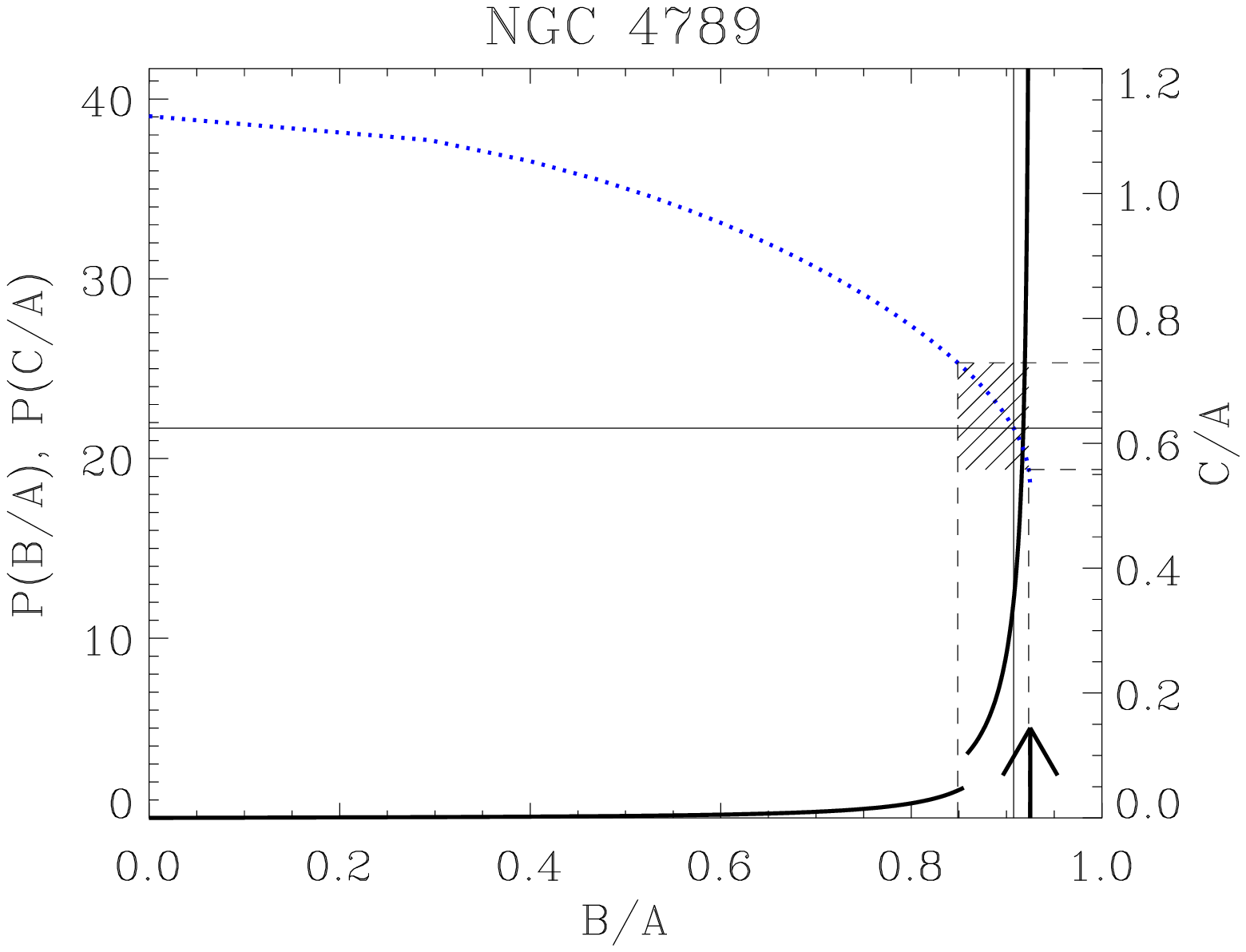}
  \caption{Relation between  the axial ratios $B/A$ and  $C/A$ for two
    sample  bulges. MCG~-02-33-017  (left panel)  hosts a  bulge with
    $\phi_{M} < \phi_C$ and NGC~4789  (right panel) hosts a bulge with
    $\phi_{M} >  \phi_C$. The probability associated  with each value
  of $B/A$  and its corresponding  value of $C/A$ (thick  solid line),
  the value of $C/A$ as a function of $B/A$ (dotted line), the maximum
  value of  the equatorial ellipticity  (arrow), the median  values of
  $B/A$ (vertical  thin solid line)  and $C/A$ (horizontal  thin solid
  line),  and the confidence  region which  encloses all  the possible
  values of $B/A$ and $C/A$ within a $67\%$ probability (hatched area)
  are shown in both panels.}
  \label{bavsca}%
  \end{figure*}

Since $B/A$ and $C/A$ are  both functions of the same variable $\phi$,
their probabilities are  equivalent (i.e., for a given  value of $B/A$
with probability  $P(B/A)$, the corresponding value  of $C/A$ obtained
by  Eq.  \ref{eqn:vartie}  has a  probability  $P(C/A)=P(B/A)$).  This
allows us  to obtain the range  of possible values of  $B/A$ and $C/A$
for an  observed bulge  and to constrain  its most  probable intrinsic
shape. An example  of the application of Eq.   \ref{eqn:vartie} to two
bulges of  our sample is  shown in Fig.  \ref{bavsca}.   The intrinsic
shape  of  bulges  for bulges  with  with  $\phi_C  < \phi_{\rm  M}  =
\frac{\pi}{4}+\frac{\phi_B}{2}$ is less  constrained, since the median
values  of $B/A$  and $C/A$  are less  representative of  their actual
values.    This  is  the   case  for   the  bulge   of  MCG~-02-33-017
(Fig. \ref{bavsca}, left panel).  On the contrary, the intrinsic shape
of bulges with $\phi_C > \phi_{\rm M}$ is better constrained.  This is
the case for the bulge of NGC~4789 (Fig. \ref{bavsca}, right panel).

\subsection{Statistics of the intrinsic shape of bulges}
\label{sec:resanddis}

We  derived the  triaxiality parameter,  as defined  by \cite{franx91}
$T=\frac{1-\left(\frac{\hat{B}}{\hat{A}}\right)^2}
{1-\left(\frac{\hat{C}}{\hat{A}}\right)^2}$, for the 115 sample bulges
with   a   well-constrained   intrinsic   shape  (i.e.,   those   with
$\phi_C>\phi_{\rm  M}$). $\hat{A}$, $\hat{B}$,  and $\hat{C}$  are the
lengths of  the longest, intermediate,  and shortest semi-axes  of the
triaxial  ellipsoid, respectively  (i.e., $\hat{A}  \geq  \hat{B} \geq
\hat{C}$).  This notation is different with respect to that we adopted
in the previous sections. Now prolate bulges do either lie on the disk
plane (and are similar to bars) or do stick out from the disk (and are
elongated perpendicularly to it). This change of notation is needed to
compare our results with those available in literature.

The triaxiality parameter  for bulges with $\phi_C >  \phi_{\rm M}$ is
characterized by a bimodal distribution with a minimum at $T=0.55$ and
two maxima at $T=0.05$  and $T=0.85$, respectively.  According to this
distribution, $65\%\pm4\%$ of the  selected bulges are oblate triaxial
(or axisymmetric) ellipsoids ($T<0.55$) and the remaining $35\%\pm4\%$
are prolate (or axisymmetric) triaxial ellipsoids ($T\ge0.55$).

We  investigated the  cause of  such  a bimodality  by separating  the
bulges  according to  their  S\'ersic index  ($n$) and  bulge-to-total
luminosity  ratio  ($B/T$).  Both  quantities  were  derived for  each
sample  bulge in Paper  I.  The  bimodality is  driven by  bulges with
S\'ersic  index $n>2$, or  alternatively, by  bulges of  galaxies with
$B/T>0.3$.   We  find that  $66\%\pm4\%$  of  bulges  with $n>2$  have
$T<0.55$. Their number decreases as $T$ increases from 0 to 0.55.  The
remaining  bulges have  $T>0.55$  and their  number  increases as  $T$
ranges from  0.55 to  1.  A similar  distribution is observed  for the
bulges of galaxies with $B/T>0.3$.   $67\%\pm4\%$ of them host a bulge
with $T<0.55$. Instead, the  distribution of the triaxiality parameter
of bulges of galaxies with $B/T\leq0.3$ is almost constant with a peak
at $T=0.05$.  This is true also for the bulges with $n\leq2$, although
to a  lesser degree.  The two subsamples  of bulges with  $n\leq2$ and
$n>2$ are different, as confirmed by a Kolmogorov-Smirnov test ($99\%$
confidence level). In particular,  the fraction of oblate axisymmetric
(or  nearly axisymmetric)  bulges ($T<0.1$)  is remarkably  higher for
$n\leq2$ ($27\%\pm4\%$)  than for $n>2$  ($14\%\pm3\%$).  The fraction
of  triaxial bulges  ($0.1 \leq  T \leq  0.9$) is  lower  for $n\leq2$
($71\%\pm5\%$) than for $n>2$ ($76\%\pm3\%$).  The fraction of prolate
axisymmetric (or nearly axisymmetric) bulges ($T>0.9$) for $n\leq2$ is
$2\%\pm2\%$, but $11\%\pm3\%$ for  $n>2$. The two subsamples of bulges
of galaxies with $B/T > 0.3$  and $B/T \leq 0.3$ are different too, as
confirmed by a Kolmogorov-Smirnov  test ($99\%$ confidence level). The
distribution  of bulges  with  $n\leq2$ and  bulges  of galaxies  with
$B/T\leq0.3$  appears  to be  the  same  at  a high  confidence  level
($>99\%$) as confirmed by a Kolmogorov-Smirnov test.
  
Bulges with $\phi_C  > \phi_{\rm M}$ can be  divided into two classes:
those  with  $n\leq2$  (or  $B/T\leq0.3$)  and those  with  $n>2$  (or
$B/T>0.3$).   About  $70\%$ of  bulges  with  $n\leq2$  are hosted  by
galaxies with  $B/T\leq0.3$.  The same  is true for bulges  with $n>2$
which are mostly hosted by  galaxies with $B/T>0.3$.  This agrees with
the correlation between $n$ and $B/T$.

\section{Conclusions}
\label{sec:conclu}

In this work,  we have developed a new method  to derive the intrinsic
shape  of  bulges. It  is  based  upon  the geometrical  relationships
between  the  observed  and  intrinsic  shapes  of  bulges  and  their
surrounding disks. We assumed that bulges are triaxial ellipsoids with
semi-axes of length $A$ and $B$  in the equatorial plane and $C$ along
the polar axis. The bulge shares the same center and polar axis of its
disk, which is circular and lies on the equatorial plane of the bulge.
The intrinsic  shape of the  bulge is recovered from  photometric data
only. They include the lengths $a$  and $b$ of the two semi-major axes
of the ellipse, corresponding to the two-dimensional projection of the
bulge, the twist  angle $\delta$ between the bulge  major axis and the
galaxy line of nodes, and the galaxy inclination $\theta$.  The method
is completely independent of the  studied class of objects, and it can
be applied whenever a triaxial ellipsoid embedded in (or embedding) an
axisymmetric component is considered.

We  analyzed  the  magnitude-limited  sample of  148  unbarred  S0--Sb
galaxies,  for  which  we  have  derived (Paper  I)  their  structural
parameters   by  a   detailed  photometric   decomposition   of  their
near-infrared surface-brightness distribution.

We derived  the triaxiality  parameter, as defined  by \cite{franx91},
for all of them. We found  that it follows a bimodal distribution with
a minimum  at $T=0.55$  and two maxima  at $T=0.05$  (corresponding to
oblate  axisymmetric or nearly  axisymmetric ellipsoids)  and $T=0.85$
(strongly prolate triaxial  ellipsoids), respectively. This bimodality
is driven  by bulges with S\'ersic  index $n > 2$  or alternatively by
bulges of  galaxies with a  bulge-to-total ratio $B/T >  0.3$.  Bulges
with $n  \leq 2$ and bulges of  galaxies with $B/T \leq  0.3$ follow a
similar distribution, which is different from that of bulges with $n >
2$ and bulges of galaxies with $B/T > 0.3$.

The different distribution of the intrinsic shapes of bulges according
to their S\'ersic  index gives further support to  the presence of two
bulge populations with  different structural properties: the classical
bulges,  which  are  characterized by  $n  >  2$  and are  similar  to
low-luminosity elliptical galaxies, and  pseudobulges, with $n \leq 2$
and  characterized by disk-like  properties.  The  correlation between
the intrinsic  shape of bulges with  $n \leq 2$ and  those in galaxies
with  $B/T \leq 0.3$  and between  bulges with  $n >  2$ and  those in
galaxies  with $B/T  > 0.3$  agrees with  the correlation  between the
bulge S\'ersic index  and bulge-to-total ratio of the  host galaxy, as
recently found by \cite{droryfisher07} and \cite{fisherdrory08}.

The observed bimodal distribution  of the triaxiality parameter can be
compared  to  the   properties  predicted  by  numerical  simulations.
\cite{cox06} studied the structure  of spheroidal remnants formed from
major  dissipationless  and dissipational  mergers  of disk  galaxies.
Dissipationless  remnants are  triaxial  with a  tendency  to be  more
prolate, whereas dissipational remnants  are triaxial and tend be much
closer to oblate.  In addition,  \cite{hopkins10} used
semi-empirical models to predict galaxy merger rates and contributions
to  bulge growth  as  functions  of merger  mass,  redshift, and  mass
ratio.  They found  that  high $B/T$  systems  tend to  form in  major
mergers, whereas low $B/T$ systems tend to form from minor mergers. In
this framework, bulges with $n \leq 2$, which shows a high fraction of
oblate axisymmetric (or nearly axisymmetric) shapes and have $B/T \leq
0.3$,  could be  the result  of dissipational  minor mergers.   A more
complex    scenario   including    both   major    dissipational   and
dissipationless  mergers  is  required   to  explain  the  variety  of
intrinsic shapes found for bulges with $n > 2$ and $B/T > 0.3$.

However,  high-resolution numerical  simulations  in a  cosmologically
motivated  framework  that  resolves  the bulge  structure  are  still
lacking.  The comparison of a  larger sample of bulges with a measured
intrinsic  shape and covering  the entire  Hubble sequence  with these
numerical experiments is the next logical step in addressing the issue
of bulge formation.



\begin{thebibliography}{}

\bibitem[1]{driver07}
{Driver}, S.~P., {Allen}, P.~D., {Liske}, J., \& {Graham}, A.~W. 2007, \apjl,
  657, L85

\bibitem[2]{daviesillingworth83}
{Davies}, R.~L. \& {Illingworth}, G. 1983, \apj, 266, 516

\bibitem[3]{corsini99}
{Corsini}, E.~M., {Pizzella}, A., {Sarzi}, M., {et~al.} 1999, \aap, 342, 671

\bibitem[4]{pignatelli01}
{Pignatelli}, E., {Corsini}, E.~M., {Vega Beltr{\'a}n}, J.~C., {et~al.} 2001,
  \mnras, 323, 188

\bibitem[5]{zaritskylo86}
{Zaritsky}, D. \& {Lo}, K.~Y. 1986, \apj, 303, 66

\bibitem[6]{bertola91}
{Bertola}, F., {Vietri}, M., \& {Zeilinger}, W.~W. 1991, \apjl, 374, L13

\bibitem[7]{mendezabreu08}
{M{\'e}ndez-Abreu}, J., {Aguerri}, J.~A.~L., {Corsini}, E.~M., \& {Simonneau},
  E. 2008, \aap, 478, 353

\bibitem[8]{gerhardvietri86}
{Gerhard}, O.~E. \& {Vietri}, M. 1986, \mnras, 223, 377

\bibitem[9]{falconbarroso06}
{Falc{\'o}n-Barroso}, J., {Bacon}, R., {Bureau}, M., {et~al.} 2006, \mnras,
  369, 529

\bibitem[10]{pizzella08}
{Pizzella}, A., {Corsini}, E.~M., {Sarzi}, M., {et~al.} 2008, \mnras, 387, 1099

\bibitem[11]{corsini03}
{Corsini}, E.~M., {Pizzella}, A., {Coccato}, L., \& {Bertola}, F. 2003, \aap,
  408, 873

\bibitem[12]{coccato04}
{Coccato}, L., {Corsini}, E.~M., {Pizzella}, A., {et~al.} 2004, \aap, 416, 507

\bibitem[13]{coccato05}
{Coccato}, L., {Corsini}, E.~M., {Pizzella}, A., \& {Bertola}, F. 2005, \aap,
  440, 107

\bibitem[14]{fathipeletier03}
{Fathi}, K. \& {Peletier}, R.~F. 2003, \aap, 407, 61

\bibitem[15]{noordermeervanderhulst07}
{Noordermeer}, E. \& {van der Hulst}, J.~M. 2007, \mnras, 376, 1480

\bibitem[16]{mosenkov10}
{Mosenkov}, A.~V., {Sotnikova}, N.~Y., \& {Reshetnikov}, V.~P. 2010, \mnras,
  401, 559

\bibitem[17]{mendezabreu10}
{M{\'e}ndez-Abreu}, J., {Simonneau}, E., {Aguerri}, J.~A.~L., \& {Corsini}, E. M. 2010, \aap, 521, 71

\bibitem[18]{simonneau98}
{Simonneau}, E., {Varela}, A.~M., \& {Munoz-Tunon}, C. 1998, Nuovo Cimento B
  Serie, 113, 927

\bibitem[19]{franx91}
{Franx}, M., {Illingworth}, G., \& {de Zeeuw}, T. 1991, \apj, 383, 112

\bibitem[20]{droryfisher07}
{Drory}, N. \& {Fisher}, D.~B. 2007, \apj, 664, 640

\bibitem[21]{fisherdrory08}
{Fisher}, D.~B. \& {Drory}, N. 2008, \aj, 136, 773

\bibitem[22]{cox06}
{Cox}, T.~J., {Dutta}, S.~N., {Di Matteo}, T., {et~al.} 2006, \apj, 650, 791

\bibitem[23]{hopkins10}
{Hopkins}, P.~F., {Bundy}, K., {Croton}, D., {et~al.} 2010, \apj, 715, 202

\end{thebibliography}
\end {document}